\documentclass{appolb}

\def\slashchar#1{\setbox0=\hbox{$#1$}
   \dimen0=\wd0 \setbox1=\hbox{/} \dimen1=\wd1
   \ifdim\dimen0>\dimen1 \rlap{\hbox to \dimen0{\hfil/\hfil}} #1
   \else  \rlap{\hbox to \dimen1{\hfil$#1$\hfil}} / \fi}

\usepackage{cite}
\usepackage{epsfig}
\usepackage{graphicx}
\usepackage{epstopdf} 
\usepackage{amsfonts}
\usepackage{dcolumn}

\def\slashchar#1{\setbox0=\hbox{$#1$}
   \dimen0=\wd0 \setbox1=\hbox{/} \dimen1=\wd1
   \ifdim\dimen0>\dimen1 \rlap{\hbox to \dimen0{\hfil/\hfil}} #1
   \else  \rlap{\hbox to \dimen1{\hfil$#1$\hfil}} / \fi}

\begin{document}

\title{
$0^{++}$ states in a large-$N_c$ Regge 
approach\footnote{Presented by ERA at Miniworkshop {\it Understanding Hadronic Spectra}, Bled (Slovenia) 3-10 July 2011}}

\author{Enrique Ruiz Arriola
\address{Departamento de F\'{\i}sica At\'omica, Molecular y Nuclear 
\\ 
and Instituto Carlos I de F{\'\i}sica Te\'orica y Computacional,
\\Universidad de Granada, E-18071 Granada, Spain}
\and 
Wojciech Broniowski
\address{The H. Niewodnicza\'nski Institute of Nuclear Physics PAN, PL-31342 Krak\'ow, Poland, and\\
Institute of Physics, Jan Kochanowski University, PL-25406~Kielce, Poland}}

\date{12 October 2011}

\maketitle
\begin{abstract}
Scalar-isoscalar states ($J^{PC}=0^{++}$) are 
discussed within the
large-$N_c$ Regge approach. We find that the lightest $f_0(600)$
scalar-isoscalar state fits very well into the pattern of the radial
Regge trajectory where the resonance nature of the states 
is advantageously used. We confirm the obtained mass values from an analysis
of the pion and nucleon spin-0 gravitational form factors, recently
measured on the lattice. 
We provide arguments suggesting an alternating meson-glueball 
pattern of the $0^{++}$ states, which is supported by the
pseudoscalar-isovector $0^{+-}$ excited spectrum and asymptotic chiral
symmetry. Finally, matching to the OPE requires a fine-tuned mass
condition of the vanishing dimension-2 condensate in the Regge
approach with infinitely many scalar-isoscalar states. 
\end{abstract} 

\PACS{12.38.Lg, 11.30, 12.38.-t}

\section{Introduction \label{sec:intro}}

The goal of this talk is to discuss various intriguing aspects of the 
spectrum of scalar-isoscalar states. Approaches developed in recent years 
may shed new light on this long-elaborated problem in hadronic physics. 

\begin{figure}[tbc]
\includegraphics[width=.47\textwidth]{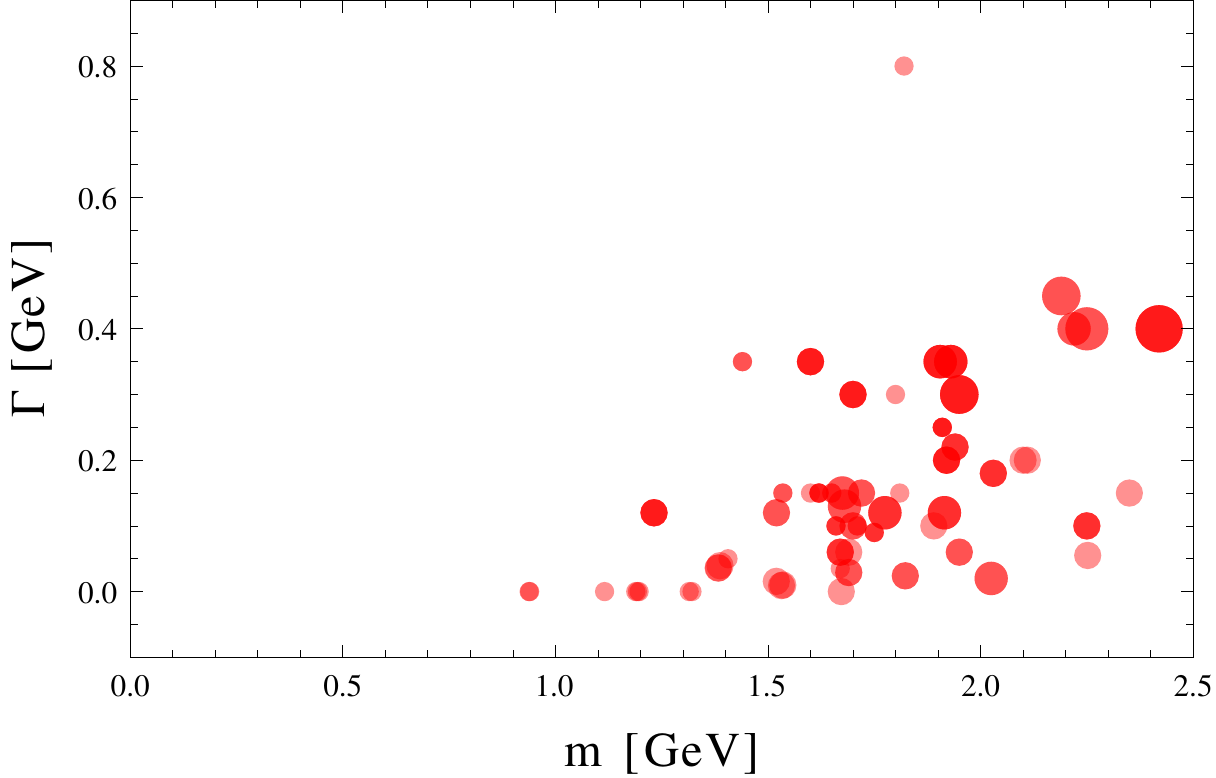}\includegraphics[width=.47\textwidth]{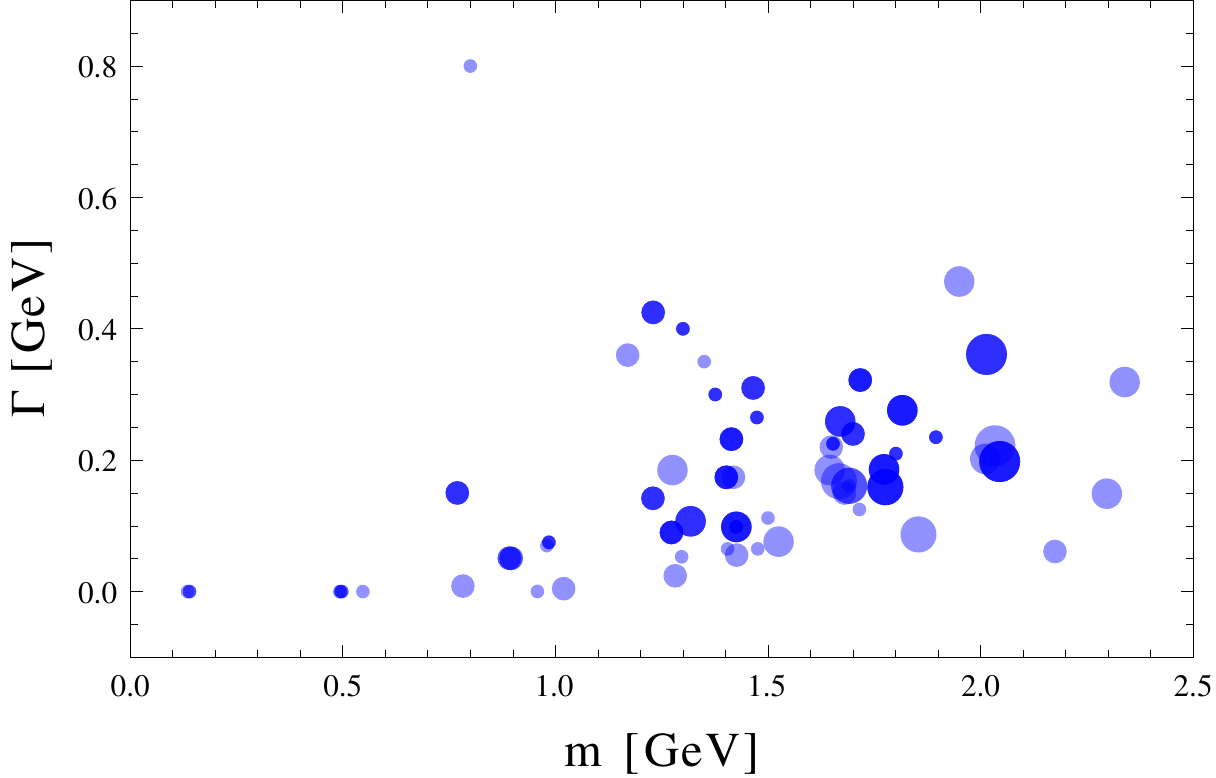}
\caption{\footnotesize Width  of {\it all} baryons (left panel) and mesons (right panel) listed in the PDG
  tables~\cite{Amsler:2008zz} as a function of the hadron mass (in
  GeV).  The surface of each point is proportional to the $(2J+1)$
  spin degeneracy, while the intensity is proportional to the isospin degeneracy $(2I+1)$.}
\label{fig:Gam-m-mes}
\end{figure}

\begin{figure}[tb]
\includegraphics[width=.47\textwidth]{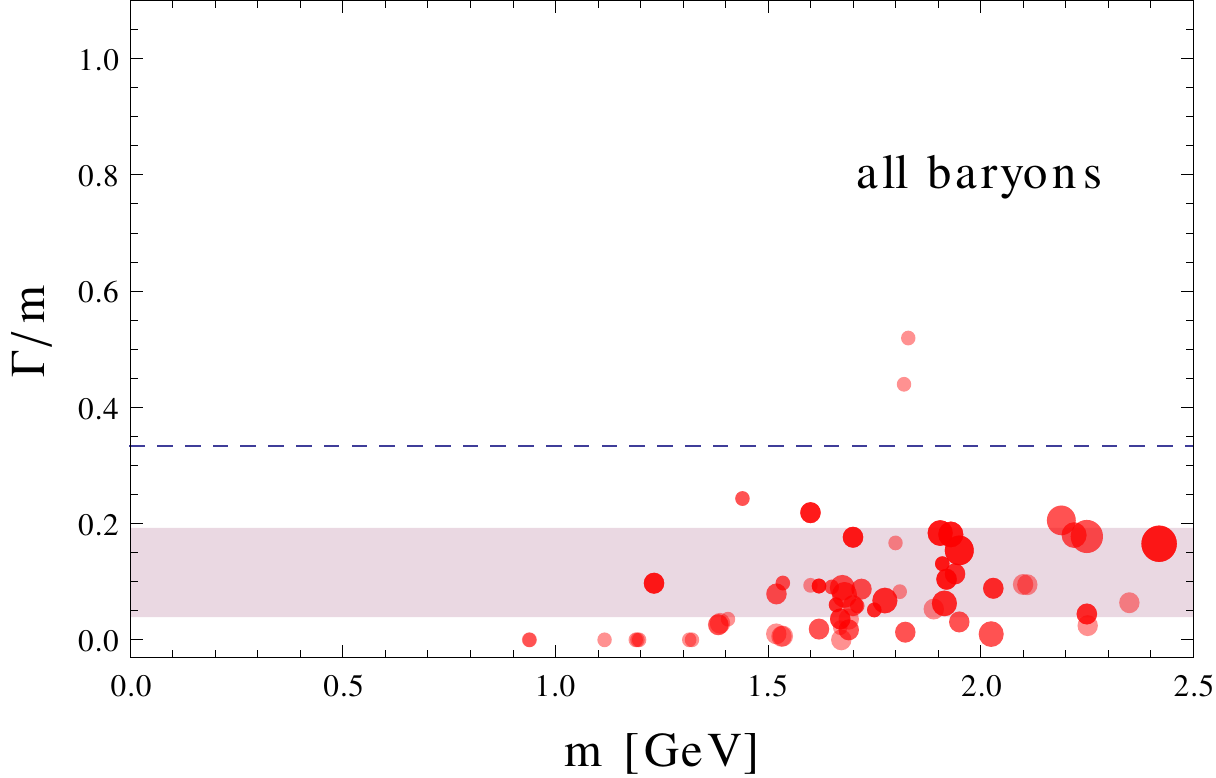}\includegraphics[width=.47\textwidth]{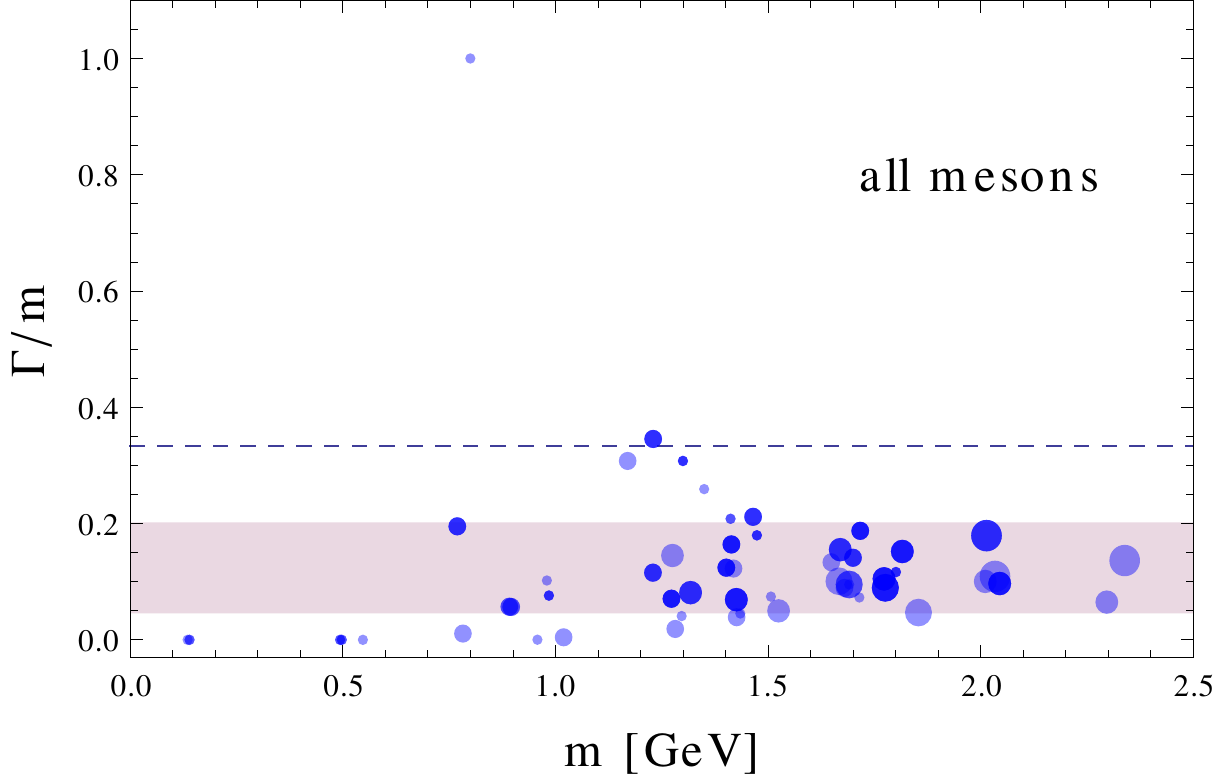}
\caption{\footnotesize Same as Fig.~\ref{fig:Gam-m-mes} for the width/mass ratio. The large-$N_c$ 
 limit predicts $\Gamma/m \sim {\cal
    O} (N_c^{-1})$.  The dashed horizontal line corresponds to 1/3.  The
    gray bands reflect the uncertainties of the fit $ \Gamma/m = 0.12(8)$.}
\label{fig:pmes}
\end{figure}

Subsequent hadron resonances listed in the Particle Data Group (PDG) tables
increase their mass up to the upper experimental limit of 2.5~GeV, while their 
width remains bound within 500~MeV. In Fig.~\ref{fig:Gam-m-mes} we show separately 
the widths of all baryons and
mesons listed in the PDG tables~\cite{Amsler:2008zz} as functions of
the mass of the state. One naturally expects that broad resonances,
i.e., with $\Gamma \sim m $, escape phenomenological analysis; even if
they existed, they might be missing from the PDG as difficult to assess 
experimentally. Note, however, that
with the exception of the notorious $f_0(600)$ resonance 
and a few baryon and meson states, 
the ratio is bound by the line $\Gamma/m \sim 1/3$ (the dashed line in Fig.~\ref{fig:pmes}). 

A natural and model-independent framework to understand this feature is provided
by the limit of large number of colors in QCD. Indeed, in this
large-$N_c$ limit, with $g^2 N_c$ fixed, baryons are heavy with mass
$m = {\cal O} (N_c)$ and width $\Gamma={\cal O} (N_c^0)$~\cite{Witten:1979kh,Broniowski:1994gm}, 
while mesons and glueballs are stable, with
mass independent of $N_c$, namely $m={\cal O}(N_c^0)$, and 
width $\Gamma$ suppressed as $1/N_c$ and $1/N_c^2$,
respectively. This means that $\Gamma/m$ is suppressed (see, e.g., \cite{Pich:2002xy} for 
a review). In particular,  one has $\Gamma/m \sim
N_c^{-1}$ for mesons and baryons, while $\Gamma/m \sim N_c^{-2}$ for
glueballs.%
\footnote{Fig.~\ref{fig:Gam-m-mes} suggests that it is
reasonable to assume that excited states in the spectrum
follow a more accurate large-$N_c$ pattern than the ground state.}

On the other hand, at high excitations mesons are expected to resemble
strings of length $l \sim m /\sigma $, with $\sigma$ denoting the string tension
(see, e.g., Ref.~\cite{Andrianov:2005aj} and references therein).
Actually, the decay rate of a string per unit time, $\Gamma$, is estimated
to be proportional to the length~\cite{Gurvich:1979nq,Casher:1978wy}, which immediately 
yields constant $\Gamma/m$.%
\footnote{The argument directly carries over to baryons, treated as a quark-diquark string.} 
In Fig.~\ref{fig:pmes}
we show the ratio of widths to masses of {\it all} baryons and mesons listed in
the PDG tables~\cite{Amsler:2008zz}, plotted as functions of the mass of the
state.  If we compute the average weighted with the
$(2J+1)$ spin degeneracy and its spread, we find ($\alpha$ are the remaining quantum numbers, including isospin) 
\begin{eqnarray}
\frac{\Gamma}{m}  \equiv \sum_{J,\alpha} (2J+1) \frac{\Gamma_{J,\alpha}}{m_{J,\alpha}} = 0.12(8)  
\label{eq:G/m}
\end{eqnarray}
for both baryons and mesons! 
This rather small ratio, complying to the large-$N_c$ and string-model arguments, 
suggests that this is a generic feature of the
hadronic spectrum rather than a lack of experimental ability to resolve too broad
states. One may thus assert in this regard that the PDG spectrum is fairly
complete.

\section{Masses of resonances \label{sec:masses}}

A resonance may be interpreted as a
superposition of states with a given mass distribution,
approximately spanning the $m \pm \Gamma/2$ interval. Of course,
the shape of the distribution depends on the particular
process where the resonance is produced, and thus on the background.
The rigorous quantum-mechanical definition of the resonance corresponds to a pole in
the second Riemann sheet in the partial-wave amplitude of the
corresponding decay channel, which becomes independent of
the background. However, although quoting the pole is highly
desirable, with a few exceptions this is {\it not} what one typically
finds in the PDG.


As a matter of fact, several
definitions are used: pole in the second Riemann sheet, pole in the $K$-matrix,
Breit-Wigner resonance, maximum in the speed plot, time delay, etc.~(see,
e.g., \cite{Suzuki:2008rp,Workman:2008iv}). Clearly, while all these
definitions converge for narrow resonances, even for broad states we
expect the masses to be compatible within their corresponding $m \pm
\Gamma/2$ intervals. As mentioned, the values listed in the PDG for a
given resonance correspond to different choices and/or processes, but
mostly the results are compatible within the estimated width differences.

The lowest resonance in QCD is the $0^{++}$ state $f_0(600)$ or the
$\sigma-$meson. It appears as a complex pole in the second Riemann
sheet of the $\pi\pi$ scattering amplitude at $\sqrt{s_\sigma} = m_\sigma -
i \Gamma_\sigma/2$ with  $m_\sigma  = 441^{+16}_{-8}$ and 
$\Gamma_\sigma = 544^{+18}_{-24} {\rm ~MeV}$~\cite{Caprini:2005zr} (see also
Ref.~\cite{GarciaMartin:2011jx}).  While these are remarkably accurate,
it is unclear whether these numbers can be directly used in
hadronic physics. An analysis of the role played by the $\sigma$ as a
correlated $2\pi$ exchange in the central component to the $NN$ force
shows that the complex-pole exchange does not accurately describe this
effect in the range $1~ {\rm fm} < r < 5~ {\rm fm}$, but prefers a value
in between the pole and the Breit-Wigner approximation,
$m_\sigma=600(50)$~MeV~\cite{CalleCordon:2009ps}. The
Breit-Wigner parameters are $(m_\sigma,\Gamma_\sigma)= (841(5),820(20))$~MeV~\cite{Nieves:2011gb}.  
For the time delay method we obtain 
$(m_\sigma,\Gamma_\sigma) \sim (475,630)$~MeV. As we see, the
different determinations agree within the wide $m_\sigma \pm \Gamma_\sigma/2$ interval.

\begin{figure}[tb]
\begin{center}
\includegraphics[width=.47\textwidth]{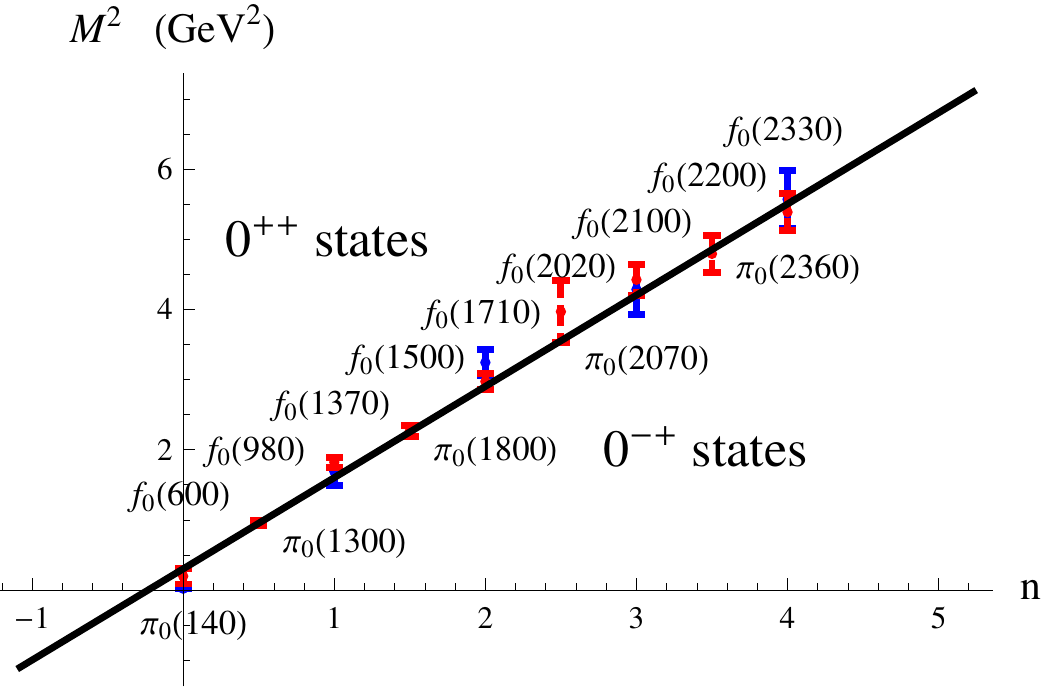}
\end{center}
\caption{\footnotesize Radial Regge trajectory corresponding to the squared mass of
  all $J^{PC}=0^{++}$ scalar-isoscalar and $J^{PC}=0^{-+}$
  pseudoscalar-isovector states listed in the PDG
  tables~\cite{Amsler:2008zz}.  The four heaviest $0^{++}$ and two
  $0^{-+}$ states are not yet well established and are omitted from
  the PDG summary tables.  The error bars correspond to the errors in
  the determination of the square of mass as $\Delta m^2 = m \Gamma $
  with $\Gamma$ from~\cite{Amsler:2008zz}. The straight line is the
  result of our joined fit. Labels of $0^{++}$ states 
are above their mark
whereas labels of $0^{+-}$ states are below their mark.}
\label{fig:regge-trajectory}
\end{figure}

\section{Scalar Regge spectrum \label{sec:regge}}

Higher $0^{++}$ states listed in the PDG~\cite{Amsler:2008zz} follow
roughly the general pattern of increasing mass but not their width.
Radial and rotational Regge trajectories were analyzed in
Ref.~\cite{Anisovich:2000kxa}. For scalar
states~\cite{Anisovich:2005jn} two parallel radial trajectories could
then be identified, including three states per trajectory. In a recent
work~\cite{RuizArriola:2010fj,Arriola:2010aj} (see
also~\cite{dePaula:2009za}) we have analyzed all the $0^{++}$ states
which appear in the PDG tables (see Fig.~\ref{fig:regge-trajectory})
and found that {\it all} fit into a single radial Regge trajectory of
the form
\begin{eqnarray}
M_S (n)^2 = \frac{a}{2} n + m_\sigma^2. \label{eq:massS}
\end{eqnarray} 
The mass of the $\sigma$ state can be deduced from this trajectory as
the mass of the lowest state. The resonance nature of these states
suggests to use the corresponding half-width as the mass uncertainty in the $\chi^2$ fit:
\begin{eqnarray}
\chi^2 = \sum_n \left(\frac{M_{f,n}-M_S(n)}{\Gamma_{f,n}/2} \right)^2 .
\label{eq:chi2}
\end{eqnarray}
Minimization yields $\chi^2 /{\rm DOF} = 0.12 $, with 
\begin{eqnarray}
a = 1.31(12) {\rm ~GeV}^2, \quad \; m_\sigma = 556(127) {\rm ~MeV}.  \label{fitII}  
\end{eqnarray} 
Formula (\ref{eq:massS}) is actually equivalent to two parallel radial Regge
trajectories with the {\em standard} slope,
\begin{eqnarray}
M_{S,-} (n)^2 &=& a \, n + m_\sigma^2,  \label{traj-1}  \\
M_{S,+} (n)^2 &=& a \, n + m_\sigma^2 + \frac{a}{2}  , \label{traj-2}
\end{eqnarray} 
where $a=2 \pi \sigma$, and $\sigma$ is the string tension associated
to the potential $V(r)= \sigma r$ between heavy colored sources.  The
value $\sqrt{\sigma} = 456(21) \, {\rm MeV}$ obtained from our fit agrees well with lattice
determinations of $\sqrt{\sigma} = 420~ {\rm MeV}$~\cite{Kaczmarek:2005ui}.  Of course, one expects some of these
states to correspond eventually to glueballs.  However, there seems to be no
obvious difference between mesons and glueballs, as far as the
radial Regge spectrum is concerned. Note that Casimir scaling suggests that the
string tension is $ \sigma_{\rm glueball}= \frac94 \sigma_{\rm meson}$, but this holds in the case of fixed and heavy sources. The
fact that we have light quarks might explain why we cannot allocate
easily the Casimir scaling pattern in the light-quark scalar-isoscalar spectrum.

\section{Interpolating fields \label{sec:fields}}

For scalar states a measure of the spectrum is given in terms of the
(gauge and renorm invariant) trace of the energy momentum tensor~\cite{Donoghue:1991qv}
\begin{eqnarray}
\Theta^\mu_\mu \equiv  \Theta  = 
\frac{\beta(\alpha)}{2\alpha}
G^{\mu\nu a} G_{\mu\nu}^a + \sum_q m_q \left[ 1 + \gamma_m (\alpha)
\right] \bar q q . 
\label{def:theta}
\end{eqnarray} 
Here $\beta(\alpha) = \mu^2 d \alpha / d\mu^2 $ denotes the beta
function, $\alpha=g^2/(4\pi)$ is the running coupling constant,
$\gamma_m(\alpha) = d \log m / d \log \mu^2 $ is the anomalous
dimension of the current quark mass $m_q$, and $G_{\mu\nu}^a$ is the
field strength tensor of the gluon field. 

It is interesting to consider the situation of massless quarks, where
things become somewhat simpler. Then, we have in principle two scalar
operators with smallest canonical
dimensions, the gluon $G^2$ and the quark $\bar q q $.
While these two operators are both scalars, they are chirally even and
odd, respectively, i.e., under the $q \to \gamma_5 q$ transformation.
Because the chiral symmetry is spontaneously broken, there is some
mixing between $G^2$ and $\bar q q $. These operators connect
scalar states to the vacuum through the matrix element
\begin{eqnarray}
 \langle 0 | \Theta | n
\rangle = m_n^2 f_n \, , \qquad  \langle 0 | \bar q q | n
\rangle = m_n c_n \, . 
\label{dM}
\end{eqnarray}
The two-point correlators read 
\begin{eqnarray}
\Pi_{\Theta \Theta} (q) &=& i \int d^4 x e^{i q \cdot x} \langle 0 |
 T \left\{ \Theta (x) \Theta (0) \right\} | 0 \rangle = \sum_n \frac{f_n^2 m_n^4}{m_n^2 -q^2} + {\rm c.t.}  \, , \\ 
\Pi_{\Theta S } (q) &=& i \int d^4 x e^{i q \cdot x} \langle 0 |
 T \left\{ \Theta (x) \bar q q (0) \right\} | 0 \rangle
 = \sum_n \frac{f_n^2 m_n^2 c_n m_n}{m_n^2 -q^2} + {\rm c.t.} \, , \\ 
 \Pi_{S S } (q) &=& i \int d^4 x e^{i q \cdot x} \langle 0 |
 T \left\{ \bar q q (x) \bar q q (0) \right\} | 0 \rangle
 = \sum_n \frac{c_n^2 m_n^2}{m_n^2 -q^2} + {\rm c.t.} \, , 
\end{eqnarray} 
where in the r.h.s. we saturate with scalar states and ${\rm c.t.} $
stands for subtraction constants which can chosen as to replace $m_n^2
\to q^2$ in the numerator. In that scheme, in the large $-q^2 \gg
\Lambda_{\rm QCD}$ limit, a comparison with the Operator Product
Expansion (OPE)~\cite{Narison:1996fm,Narison:1984bv,Harnett:2008cw} leads to the matching conditions
\begin{eqnarray} 
\Pi_{\Theta \Theta } (q^2) &=& q^4 \,  C_{0} \, \log (-q^2) + \dots,  \nonumber \\ 
\Pi_{S S }(q^2 ) &=&   q^2 \,  C_{0}'  \,  \log (-q^2) +  \dots,   \nonumber \\
\Pi_{\Theta S }(q^2 ) &=&  C_{0}'' \, \langle \bar q q  \rangle \log (-q^2) + \dots, 
\label{qcdsr}
\end{eqnarray} 
where $C_0=-(2 \beta(\alpha)/\alpha \pi)^2 $, $C_0'=-3/( 8\pi^2)$ and $C_0''=-2 \beta(\alpha)/\alpha \pi$. As we see, $\bar q q $ and
$G^2$ do not mix at high $q^2$ values, a consequence of asymptotic chiral
symmetry~\cite{Donoghue:1991qv}. In these limits the sums over $n$ can
be replaced by integrals, whence the following asymptotic conditions
are found:
\begin{eqnarray} 
f_n^2 / (d  m_n^2 /dn ) \to C_0 \, , \quad 
c_n^2 / (d  m_n^2 /dn ) \to C_0' \, , \quad 
c_n f_n m_n^3 / (d  m_n^2 /dn ) \to C_0'' \langle \bar q q \rangle \, .   
\end{eqnarray}  
We see that the first two conditions are incompatible with the third
one if $m_n $ increases for large $n$, as is the case of the data.  
However, if we group the states in two families, as suggested by Eqs.~(\ref{traj-2}) and 
Fig.~\ref{fig:regge-trajectory}, we get a compatible solution 
\begin{eqnarray}
c_{n,-} , f_{n,+}\to {\rm const} \qquad c_{n,+}, f_{n,-} \to {\rm const}/m_n^3.
\label{eq:asympto}
\end{eqnarray}
This is equivalent to assuming an asymptotically alternating pattern
of mesons and glueballs, coupling to chirally odd and even operators,
$\bar q q $ and $G^2$, respectively. Since asymptotically $m_n^2 \sim
a n/2$, we find $c_{n,-}/c_{n,+}$ and $f_{n,+}/f_{n,-} \sim
n^\frac32$. Of course, this is not the only solution.

The situation described above suggests the existence
of a hidden symmetry in the $0^{++}$ sector. In our case 
we could think of the $\gamma_5$-parity (which becomes a good quantum number for excited
states) as the relevant symmetry which makes the difference
between the chirally even and odd states. This, however, only explains
the fact that asymptotically the slopes of the $+$ and $-$ branches are the same, but not why 
the intercepts accurately differ by half the slope.

\section{The holographic connection \label{sec:holography}}

To further elaborate on this intriguing point of the accidental degeneracy, let us consider the
one-dimensional harmonic oscillator with frequency $\omega$, as an
example; all states $\psi_n(z)$ with the energy $E_n = \hbar \omega
(n+1/2)$ can be separated into parity {\it even} and parity {\it odd }
states, satisfying the conditions $\psi_{n,\pm} (z) = \psi_{2n}(z)$ and $\psi_{\pm,n} (-z) = \pm
\psi_{\pm,n}(z)$, respectively, and having the energies $E_{+,n} = 2 \hbar \omega ( n+1/4)$ and
$E_{-,n} = 2 \hbar \omega ( n+3/4)$. These formulas display {\it twice}
the slope of $E_n$. Thus, given the states with energies $E_{+,n}$ and
$E_{-,n}$, we might infer that parity was a hidden symmetry of a
Hamiltonian explaining the correlation between the slope and intercepts.

In the relativistic case the argument can also be made in a suggestive
manner. Let us consider the Klein-Gordon action for
infinitely many bosons in four dimensions, described with fields $\phi_n (x)$ of masses $m_n$:
\begin{eqnarray}
S = \frac12 \int d^4 x \sum_{n} \left[ \partial^\mu \phi_n  \partial_\mu \phi_n  
-m_n^2 \phi_n^2 \right] .
\end{eqnarray}
We assume the spectrum of the form $m_n^2 = a n + m_0^2$.  Next, we
can introduce the five-dimensional field $\phi(x,z) = \sum_n \phi_n
(x) \psi_n(z)$, with $\psi_n(z)$ fulfilling the auxiliary
Sturm-Liouville problem in the variable $0 \le z < \infty $,
\begin{eqnarray}
-\partial_z \left[ p(z) \partial_z \psi_n(z) \right] +
q(z) \psi_n(z) = m_n^2 \rho(z) \psi_n(z), 
\label{eq:sl}
\end{eqnarray}
where the functions are orthogonal with respect to the weight function
$\rho(z)$, provided suitable boundary conditions 
\begin{eqnarray}
p(z)\left(\psi_n' (z) \psi_m (z)-\psi_n (z)\psi_m'(z) \right)|_{z=0}=0
\label{eq:bc-self}
\end{eqnarray}
and $\psi_n (\infty) \to 0$ are fulfilled. 
The action can then be written as
\begin{eqnarray}
S = \frac12 \int d^4 x \int_0^\infty d z  \left[ \rho(z) \partial^\mu \phi
 \partial_\mu \phi - p(z) (\partial_z \phi)^2- q(z) 
 \phi^2 \right] 
\end{eqnarray}
after some integration by parts in the variable $z$. This action can be
written as a five-dimensional action with a non-trivial metric~\cite{Afonin:2010fr},
featuring the AdS/CFT
(soft-wall) approach (see \cite{Erdmenger:2007cm} and references
therein), with the extra dimension $z$ playing the role of a
holographic variable and the orthogonal set of functions $\psi_n(z)$
denoting the corresponding Kaluza-Klein modes.  Clearly, $z$ has the dimension
of length, suggesting that $z \to 0 $ corresponds to the ultraviolet
and $z \to \infty $ to the infrared regime. 

Turning to Eq.~(\ref{eq:sl}), we may take 
the standard Harmonic oscillator 
Schr\"odinger-like equation ($p(z)=\rho(z)=1$, $q(z) \equiv U(z) =a^2 z^2/16$) 
\begin{eqnarray}
-\psi_n''(z) + \frac1{16} a^2 z^2 \psi_n(z) = m_n^2 \psi_n(z)
\label{eq:ho}
\end{eqnarray}
and  obtain for the regular solutions at infinity the result  
\begin{eqnarray}
\frac{\psi_n'(0)}{\psi_n(0)} = - \sqrt{a} \, \frac{\Gamma\left( \frac34 - \frac{m_n^2}{a}\right)}{\Gamma\left( \frac14 - \frac{m_n^2}{a}\right)},
\label{eq:bc-log}
\end{eqnarray}
where $\Gamma(x)$ is the Euler Gamma function, which is meromorphic and 
have simple poles at $x=0,-1,-2, \dots$.  
The solutions fulfilling the Dirichlet, $\psi_n (0)=0$, and Neumann, $\psi_n'(0)=0$,
boundary conditions, respectively, have the masses
\begin{eqnarray}
m_{-,n}^2 = a n + \frac{a}{4} \, , \qquad m_{+,n}^2 = a n + \frac{3 a}{4} \, ,  
\end{eqnarray}
which can be merged into one single formula 
\begin{eqnarray}
m_n^2 = \frac{a}{4} (2 n+1) \, .
\label{eq:all_mass}
\end{eqnarray}
This yields $m_\sigma= m_{f_0}/\sqrt{3}= 566~{\rm MeV}$ and, for
the string tension, $\sigma =
m_{f_0} \sqrt{2/3\pi}=450~{\rm MeV}$ with $m_{f_0}= 980~{\rm MeV}$,
quite reasonable values. 

In this approach the symmetry in the scalar spectrum corresponds to a
parity symmetry in the holographic $z$ variable $\psi_n(-z) = \pm
\psi_n(z) $. Note that usually the holographic variable $z$ is taken
to be positive\footnote{This is supported by the light-front
  interpretation of Brodsky and de Teramond~\cite{deTeramond:2008ht},
  where the holographic variable is the polar coordinate of a two
  dimensional vector, $z = |\vec \zeta|$ and $\vec \zeta = \vec b
  \sqrt{x (1-x)}$, with $\vec b$ denoting the impact parameter and $x$ the
  momentum fraction of the quark. This interpretation yields a two
  dimensional potential $U(\vec \zeta)= \kappa^2 \vec \zeta^2  + 2
  \kappa^2 (L+S-1)$ with $J=L+S$ which, when passing to the polar
  variable $z$, generates the usual centrifugal term $(L^2-1/4)/z^2$
  not present in our discussion, yielding $M_{n,L,S}^2 = 4 \kappa^2 (n+L+S/2)$
which for $J=0$ and $L=1$ resembles Eq.~(\ref{eq:all_mass}).}, but if we extend it to $-\infty < z
< \infty $, we may define a holographic superfield containing two
different and orthogonal modes. Otherwise, in the interval $ 0 < z <
\infty$ the Dirichlet and Neumann modes are not orthogonal to each
other.

\section{Pseudoscalar mesons and chiral symmetry \label{sec:chiral}}

Discerning the nature of the $\sigma$ state has been a recurrent
pastime for many years. As is well known, glueballs are more weakly
coupled to mesons, ${\cal O} (1/N_C)$, than other mesons, ${\cal O}
(1/\sqrt{N_C})$. The minimum number of states, allowed by certain sum
rules and low energy theorems, is just two. In
Ref.~\cite{RuizArriola:2010fj} we undertake such an analysis which
suggests that $f_0(600)$ (denoted as $\sigma$) is a $\bar q q $ meson,
while $f_0(980)$ (denoted as $f_0)$ is a glueball.  This is supported
by the rather small width ratio, which yields
$\Gamma_{f}/\Gamma_{\sigma} \sim (g_{f \pi\pi}^2 m_f^3)/(g_{\sigma
  \pi\pi}^2 m_\sigma^3) \sim 1/N_C$, thus for $m_\sigma \sim 0.8~{\rm MeV} \sim m_f$ the
ratio $g_{\sigma \pi\pi}/g_{f \pi\pi} \sim \sqrt{N_C}$ is obtained.

A further piece of evidence for the alternating meson-glueball pattern
is provided by looking at the excited pion spectrum, which we show in
Fig.~\ref{fig:regge-trajectory}. The alternating pattern was unveiled
by Glozman~\cite{Glozman:2003az}, suggesting that states degenerate
with the pion might not be identified with glueballs. Remarkably, the
states generating doublets with pion states are $f_0(600)
\leftrightarrow \pi_0 (140)$, $f_0(1370) \leftrightarrow \pi_0
(1300)$, $f_0(1710) \leftrightarrow \pi_0 (1800)$,
$f_0(2100)\leftrightarrow \pi_0 (2070) $, and $f_0(2330)
\leftrightarrow \pi_0 (2360)$, whereas the other scalar states
$f_0(980)$, $f_0(1500)$, $f_0(2020)$ and $f_0(2200)$ are not
degenerate with other mesons with light $u$ and $d$ quarks.  Our
analysis is reinforced by this observation.

{As a matter of fact, fitting the pion
  $\pi(140)$ as the $n=0$ state of the Regge spectrum requires strong
  departure from a simple linear trajectory, $m_n^2 = a n +
  m_0^2$. One may improve on this by using the holographic connection
  and a mixed boundary condition at $z=0$ determined by fixing the
  mass of the ground state $m_0$ using Eq.~(\ref{eq:bc-self}) together
  with Eq.~(\ref{eq:bc-log}) for the harmonic oscillator case,
  Eq.~(\ref{eq:ho}). This procedure ensures the orthogonality between
  all states and implements linearity for large $n$. This can be done
  for the ground states $m_0=m_\pi, m_\sigma, m_{f_0}$. The fit to all
  states yields $a=1.37 {\rm GeV}^2$ and the mass spectra (in GeV)
\begin{eqnarray}
\pi\,  ({\rm Regge})\, (0.140,1.260,1.730,2.092,2.400, \dots) \,&& \pi \, ({\rm PDG})\,  (0.140,1.300,1.812,2.070,2.360)\nonumber  \\ 
\sigma \, ({\rm Regge}) \, (0.527,1.297,1.750,2.106,2.411, \dots)&&  \sigma \, ({\rm PDG})\, (0.600,1.350,1.724,2.103,2.321) \nonumber \\ 
f_0 ({\rm Regge})\,\,   (0.977,1.513, 1.906, 2.232,2.517, \dots) && f_0 ({\rm PDG})\,  (0.980,1.505,1.992,2.189)\nonumber 
\end{eqnarray}
yielding $ 1/\sqrt{a} \, \psi_0'(0)/\psi_0(0)=-3.1$, $-14.9$, and
$0.2$, respectively. Note the large and small values for the
$\sigma$ and $f_0$ cases, which suggests that these boundary conditions are
very close to the Dirichlet and Neumann cases. Chiral symmetry breaking
corresponds to the different $\pi$ and $\sigma$ values.}

\section{Scalar dominance and heavy pions \label{sec:heavy}}

Hadronic matrix elements of the energy-momentum tensor, the so-called
gravitational form factors (GFF) of the pion and nucleon, correspond
to a dominance of scalar states in the large-$N_c$ picture, as ($u(p)$
is a Dirac spinor)
\begin{eqnarray}
\langle \pi (p') | \Theta | \pi(p) \rangle &=& 
\sum_n \frac{g_{n \pi \pi} f_n q^2
m_n^2}{m_n^2-q^2}, \\ 
\langle N(p') | \Theta | N(p) \rangle &=& \bar u(p') u(p) \sum_n  \frac{   g_{n NN} f_n  m_n^2}{m_n^2-q^2},
\end{eqnarray} 
where the sum rules $\sum_n {g_{n\pi \pi}}{f_n}
=1$~\cite{Narison:1988ts} and  $M_N = \sum_n g_{n NN}
f_n$~\cite{Carruthers:1971vz} hold. Unfortunately, the lattice QCD
data for the pion~\cite{Brommel:2007xd} and nucleon (LHPC
~\cite{Hagler:2007xi} and QCDSF~\cite{Gockeler:2003jfa}
collaborations), picking the valence quark contribution, 
are too noisy as to pin down the coupling of the excited scalar-isoscalar states to
the energy-momentum tensor.  Nevertheless, useful information
confirming the (Regge) mass estimates for the $\sigma$-meson can be
extracted~\cite{RuizArriola:2010fj} through the use of the multiplicative QCD evolution
of the GFF in the valence quark momentum fraction, $\langle x
\rangle_{u+d}$, as seen in deep inelastic scattering or on the lattice
at the scale $\mu= 2~{\rm GeV}$. For the pion and nucleon GFFs we
obtain the fits
\begin{eqnarray}
\langle x \rangle_{u+d}^\pi &=& 0.52(2), \qquad \quad m_\sigma = 445(32) {\rm ~MeV} \, ,  \label{opti2} \\ 
\langle x \rangle_{u+d}^N &=& 0.447(14), \qquad \ m_\sigma = 550^{+180}_{-200} {\rm MeV}.
\end{eqnarray}
Assuming a simple dependence of $m_\sigma$ on $m_\pi$,
\begin{eqnarray}
m_\sigma^2(m_\pi)=m_\sigma^2+c\left  ( m_\pi^2-m_{\pi,{\rm phys}}^ 2 \right ), \label{msmpi}
\end{eqnarray}
yields $ m_\sigma = 550^{+180}_{-200} {\rm MeV}$ and $
c=0.95^{+0.80}_{-0.75}$, or $\ m_\sigma = 600^{+80}_{-100} {\rm MeV} $
and $ c=0.8(2)$, depending on the choice of the lattice data~\cite{Hagler:2007xi}
or~\cite{Gockeler:2003jfa}, respectively. Note that $c$ is close to
unity.  {\it Higher} quark masses might possibly clarify whether or
not the state evolves into a glueball or a meson. For a $(\bar q q
)^n$ system one expects $m_\sigma \to 2 n m_q + {\rm const}$ at large
current quark mass $m_q$.  The data from~\cite{Hagler:2007xi}
or~\cite{Gockeler:2003jfa} are too noisy to see the difference,
although for the largest pion masses we see that $m_\sigma \sim m_\pi$,
as it simply corresponds to the $\bar q q $-component of $\Theta$.  We observe,
however, that for $m_\pi \sim 500 {\rm MeV}$ our results are not far away from
the recent lattice calculation using the tetraquark probes, $(\bar q q
)^2$~\cite{Prelovsek:2010kg}, which provide $m_\sigma \sim 2 m_\pi$ for
the largest pion masses as they should (see
Table~\ref{tab:mass-fits}). From this viewpoint, and unless operator
mixing is implemented, the nature of the state is predetermined by the
probing operator.

\begin{table}[tb]
\begin{center}
\begin{tabular}{lrrrr}
\hline 
{$m_\pi$[MeV]}
& {$m_\sigma$ [MeV]} 
& {$m_\sigma$ [MeV]}
& {$m_\sigma$ [MeV]} 
& {$m_\sigma$ [MeV]} \\ 
\hline 
  
& {GFF} 
& {GFF}
& {$(\bar q q)^2$-dynam. } 
& {$(\bar q q)^2$-quench.} \\ 
\hline 
230  & 580(190)     &  620(100) & --  & 400(30)        \\ 
342  & 630(190)     &  660(90) & --  & 720(20)        \\ 
478  & 710(200)     &  730(90) & --  & 1000(20)      \\ 
318  & 620(190)     &  650(90) &  468(50)  &   --    \\ 
469  & 700(190)     &  720(90) &  936(13) &    --     \\ 
526  & 739(200)     &  750(90) &  1066(13)  &  --       \\
\hline
\end{tabular}
\end{center}
\caption{\label{tab:mass-fits} Scalar monopole mass obtained from the
  nucleon gravitational form factors, extracted from the $(\bar q q)$
  components obtained by LHPC~\cite{Hagler:2007xi} and
  QCDSF~\cite{Gockeler:2003jfa} and compared to the lattice
  calculation using the tetraquark $(\bar q q)^2$ probing fields, both
  for the dynamical and quenched fermions~\cite{Prelovsek:2010kg} .}
\end{table}

\section{Dimension-2 condensates \label{sec:dim2}}

One of the problems of the large-$N_c$ Regge models~\cite{RuizArriola:2006gq} and their holographic
relatives~\cite{Andreev:2006vy,Zuo:2008re} is that they may contradict expectations from the OPE, as
they involve dimension-2 condensates. For instance, the OPE for the
$\Pi_{\Theta \Theta} (q^2)$ correlator in Eq.(\ref{qcdsr}) gives
corrections ${\cal O} (q^0)$, while the ${\cal O} (q^2)$ terms are
missing~\cite{Narison:1996fm}. This yields a one to one comparison:
\begin{eqnarray}
C_0 &=& 
-\lim_{n \to \infty }\,  \frac{f_n^2}{d  m_n^2 /dn }
=  - \frac{ N_c^2-1}{2\pi^2} \left(\frac{\beta(\alpha)}{\alpha}\right)^2
 \label{eq:C0},  \\ 
C_2 &=&  \sum_n f_n^2  \qquad \qquad = 0 \,   \label{eq:C2},  \\ 
C_4 &=&  \sum_n f_n^2 m_n^2 \qquad \quad = \left(\frac{\beta(\alpha)}{\alpha}\right)^2 \langle G^2 \rangle \,. 
\end{eqnarray} 
Equation~(\ref{eq:C0}) requires infinitely many states, while
Eq.~(\ref{eq:C2}) suggests a positive and non-vanishing
gauge-invariant dimension-2 object, $C_2= i \int d^4 x x^2 \langle
\Theta (x) \Theta \rangle $, which is generally non-local, as it
should not appear in the OPE. Note that $C_2 > 0$, hence is
non-vanishing for a finite number of states.  The infinite Regge
spectrum of Eq.~ (\ref{eq:massS}) with Eq.~(\ref{eq:C0}) may be
modeled with a constant $f_{f_0}=f_{n,+}={\cal O}(N_c)$ whereas 
$f_{n,-}={\cal O} (\sqrt{N_c})$ goes as Eq.~(\ref{eq:asympto}) 
and yields a convergent and hence positive contribution. Naively, we get
$C_2=\infty$. However, $C_2$ may vanish, as required by standard OPE,
when infinitely many states are considered {\it after
  regularization}. The use of the $\zeta$-function
regularization~\cite{Arriola:2006sv,RuizArriola:2010fj} gives 
\begin{eqnarray}
C_2 \equiv  \lim_{s\to 0} \sum f_{n}^2 M_{S} (n)^{2s} = 
\sum_n f_{n_,-}^2 + f_{f_0}^2 \left( 1/2 - m_{f_0}^2/a \right) \, . \label{match:dim2}  
\end{eqnarray}
Then  $C_2=0$ for $m_{f_0} > \sqrt{a/2}= 810(40)~{\rm MeV} $, a
reasonable value to ${\cal O}(1/N_c)$
  In any case, the important
remark is that while at the OPE level $C_2=0$ vanishes for trivial
reasons, at the Regge spectrum level some fine tuning must be at work.

\vskip 1cm

{\em This work is partially supported by the Polish Ministry of Science and
Higher Education (grants N~N202~263438 and N~N202~249235), Spanish
DGI and FEDER funds (grant FIS2008-01143/FIS), Junta de Andaluc{\'\i}a
(grant FQM225-05).}

\end{document}